# Thermoelectric Performance of 2D Tellurium with Accumulation Contacts

Gang Qiu[1,2], Shouyuan Huang[2,3], Mauricio Segovia[2,3], Prabhu K. Venuthurumilli[2,3], Yixiu Wang[4], Wenzhuo Wu[4], Xianfan Xu[1,2,3★], Peide D. Ye[1,2★]

[1]School of Electrical and Computer Engineering, Purdue University, West Lafayette, Indiana 47907, USA

[2]Birck Nanotechnology Center, Purdue University, West Lafayette, Indiana 47907, USA

[3]School of Mechanical Engineering, Purdue University, West Lafayette, Indiana 47907, USA

[4]School of Industrial Engineering, Purdue University, West Lafayette, Indiana 47907, USA

★e-mail: Correspondence and requests for materials should be addressed to X.X (xxu@ecn.purdue.edu) or P. D. Y. (yep@purdue.edu)

**Abstract**

Tellurium (Te) is an intrinsically p-type doped narrow bandgap semiconductor with excellent electrical conductivity and low thermal conductivity. Bulk trigonal Te has been theoretically predicted and experimentally demonstrated to be an outstanding thermoelectric material with high value of thermoelectric figure-of-merit ZT. In view of the recent progress in developing synthesis route of two-dimensional (2D) tellurium thin films as well as the growing trend of exploiting nanostructures as thermoelectric devices, here for the first time we report excellent thermoelectric performance of tellurium nanofilms, with room temperature power factor of 31.7 μW/cm·K$^2$ and ZT value of 0.63. To further enhance the efficiency of harvesting thermoelectric power in nanofilm devices, thermoelectrical current mapping was performed with a laser as a heating source, and we found high work function metals such as palladium can form rare accumulation-type metal-to-semiconductor contacts to 2D Te, which allows thermoelectrically generated carriers to be collected more efficiently. High-performance thermoelectric 2D Te devices have broad applications as energy harvesting devices or nanoscale Peltier coolers in microsystems.

Thermoelectricity emerges as one of the most promising solutions to the energy crisis we are facing in 21st century. It generates electricity by harvesting thermal energy from ambient or wasted heat, which is a sustainable and environmental-friendly route compared to consuming fossil fuels[1,2]. The efficiency of converting heat to electricity is evaluated by the key thermoelectrical figure of merit: $ZT = S^2 \frac{\sigma}{\kappa} T$, where $S$ is the Seebeck coefficient defined as $S = \frac{\Delta V}{\Delta T}$, $\sigma$ and $\kappa$ are electrical and thermal conductivity, $V$ is the measured thermal voltage, and $T$ is the operating temperature. However the ZT value has not been significantly enhanced since 1960's[3] and so far the most state-of-the-art bulk materials can merely surpass 1 at room temperature[4,5]. This is because the parameters in defining ZT, Seebeck coefficient, electrical conductivity and thermal conductivity are usually correlated through the Wiedemann–Franz law[6–8] and by engineering one parameter, generally other parameters will compensate the change, which poses a dilemma for drastically improving ZT.

For the past decades enormous efforts have been made to increase thermoelectric efficiency along two major pathways: either by developing new high-efficiency thermoelectric bulk materials, or by developing novel nano-structured thermoelectric materials[9]. From material perspective, the paradigm of an excellent thermoelectric material should be a heavily doped narrow-bandgap semiconductor with good conductivity meanwhile because of the existence of a finite bandgap, the separation of electrons and holes can avoid opposite contributions to the Seebeck coefficient. Also, heavy elements are preferred for thermoelectrical applications since they can enhance the ZT values by providing more effective phonon scattering centers and reducing the thermal conductivity[1,10]. Furthermore, valley engineering is also proposed recently[11], where higher valley degeneracy can provide multiple carrier conducting channels without affecting the Seebeck coefficient and thermal conductivity. Interestingly, 2D Te has high degeneracy in both conduction and valence bands[12].

Structurally engineered low-dimensional nanostructure is another emerging field for thermoelectric research. The first advantage of low-dimensional nanostructure is the boundaries of the crystal will scatter phonons more effectively than electrons due to the different scattering length or mean free path of these two species. The disruptive lattice period in low-dimensional nanostructures can block the acoustic phonon propagating whereas electrons can be transmitted less interrupted. Therefore, the thermal conductivity of the nanostructure material system is reduced without significantly degrading electrical conductivity[2,9,13]. Moreover, when the characteristic length of the nanostructure geometry is small enough to influence the band structure because of the quantum confinement effect, the profile of density of states (DOS) can evolve into sharp shapes at band edges, which boosts the Seebeck coefficient, since the Seebeck coefficient is related with how fast DOS changes near the Fermi energy[9,14].

Trigonal tellurium is a simple but ideal material system for thermoelectric application which meets almost all the aforementioned criteria of a good thermoelectric material. Bulk Te has been theoretically predicted[15] and experimentally demonstrated[16] with high thermoelectric performance because it is a heavily doped narrow band gap semiconductor with good electrical conductivity. On the other hand, Te is the heaviest stable (i.e. non-radioactive) non-metallic element (excluding noble gases), therefore it has very low thermal conductivity due to the heavy atom mass, which also contributes to higher thermoelectrical efficiency. Recently, with more comprehensive band structure of Te being revealed[17,18], it has been revisited for a new generation of thermoelectric materials[19–21] due to its strong spin-orbit coupling which leads to a band splitting and quadruple band degeneracy near the edge of valence band[12].

Here, utilizing intrinsic high thermoelectric performance of tellurium combining with the advantage of nano-structure in 2D form, we report highly-efficient thermoelectric devices fabricated from two-dimensional Te nanofilms. A recently proposed and demonstrated liquid phase synthesis approach allows us to acquire large-scale ultrathin 2D Te nanofilms with high crystal quality[22,23], which enables the exploration of its application for electronics[22,24,25], piezo-electronics[26], and optoelectronics[27]. In this paper, we focus on investigating the thermoelectric performance of tellurium nanofilm devices.

**Results**

Figure 1(a) shows the atomic structure of the trigonal Te lattices. Covalently bonded atomic helical chains stretching in one direction are packed in parallel into hexagonal lattices. Each atomic chain is weakly bonded to six nearest neighboring chains, therefore tellurium can be considered as one-dimensional (1D) material in atomic structure, as an analogy to 2D materials such as graphene, molybdenum disulfide ($MoS_2$) and black phosphorus. Bulk Te has a narrow direct band gap (~0.35 eV) at H point of the Brillouin zone, with a camelback feature at the vicinity of valence band maxima[14,22] (see Figure 1(b)). With recently proposed and demonstrated solution-based hydrothermal process[22,23], large-area high-quality 2D Te films can be grown with thickness ranging from monolayer to tens of nanometers. Figure 1(c) shows typical as-grown Te nanofilms dispensed onto a Si wafer with 300 nm $SiO_2$ capping layer, with spiral atomic chains laying in parallel along the longer edge of the films, as confirmed by TEM[22] and angle-resolved Raman spectroscopy method[23]. The crystal orientation suggests these Te films, despite its 1D atomic structure nature, resemble 2D van der Waals material, such as graphene, with dangling bond-free interfaces, which allow us to explore their unique thermoelectric performance.

To measure the Seebeck coefficient and conductivity of 2D Te, thermoelectric measurement device was designed and fabricated, as shown in Figure 2(a) with a 31-nm-thick 2D flake in an optical image, similar to the measurement setup for other low-dimensional materials such as graphene[28,29], $MoS_2$[30] and $SnS_2$[31]. Two four-terminal micro-thermometers (Pin 3-6 and Pin 9-12) were placed at two ends of a Te flake along atom chain direction (as denoted in Figure 2(a)) to measure the local temperature as well as the voltage drop simultaneously. The thermometer filaments were made of 100-nm-thick platinum (Pt) nano-strip because Pt shows good linearity of temperature coefficient for broad temperature region including room temperature. The temperature versus resistance curve was calibrated and plotted in Supplementary Information Figure S1. 10 nm nickel was deposited underneath Pt for better adhesion. A microheater (Pin 1 and Pin 2) was placed near one side of the flake to generate temperature gradient. The microheater was designed intentionally at least one order longer (>200 μm) than the spacing and the flake size so that 1D thermal transport approximation is applicable. Two additional contacts (Pin 7 and Pin 8) were placed on Te as probes to measure the electrical conductivity of Te with standard four-terminal configuration. By applying certain voltage onto the microheater (between Pin 1 and 2), the heating power was added to the system. The voltage drop was measured across the flake between Pin 3 and Pin 9 after sufficiently long time until the system reaches equilibrium and the voltage was settled. The resistance of the thermometer was also measured with four-terminal configuration and was converted into local temperature using calibrated temperature vs. resistance curve (Figure S1). We observed monotonic increase of voltage drop and temperature gradient as we raised the output power (Figure 2(b)), and the temperature rise at thermometer 1 was slightly higher compared to thermometer 2 since it is closer to the heating source. The voltage drop ΔV is plotted against the temperature difference across the flake ΔT (see Figure 2(c)) with a linear trend from whose slope we can extract the Seebeck coefficient to be S = ΔV/ΔT = $413\ \mu V/K$. Furthermore, the conductivity of Te films can also be measured with four-terminal configuration using Pin 1, 7, 8 and 9. The measured Seebeck coefficient and conductivity gives a high value of power factor, which takes the form: PF = $S^2\sigma = 31.7 \mu W/cm \cdot K^2$. The thermal conductivity along the 1D chain direction of a 35-nm-thick Te film, measured using the micro-Raman spectroscopy technique[32] on the suspended 2D Te film, is about 1.50 W/m·K (also see Supplementary Note 5). This thermal conductivity value is lower than that of bulk Te ~ 3.0 W/m·K[33]. It is reasonable by considering phonon scattering at the surfaces which reduces thermal conductivity of 2D films. Based on the measured thermal conductivity of Te films, we can derive a room temperature ZT value of 0.63, indicating 2D Te possesses one of the best thermoelectric performance at room temperature, higher than the well-known thermoelectric element material such as Bi (ZT = 0.15) and Sb (ZT = 0.07).

To investigate the thermoelectric power generated across the 2D Te film, laser induced thermoelectrical (LITE) current mapping was carried out on a simple two-terminal device. In the next section we will discuss how different kinds of metal contacts will affect the efficiency of harvesting thermoelectric power and using the LITE current mapping technique to visualize the contacts at depletion-mode, neutral-mode at flatband condition and accumulation-mode.

The experiment setup is shown schematically in Figure 3(a). A He-Ne laser with a wavelength of 633 nm was used to locally heat the flake and create a temperature gradient. The short-circuit thermoelectric current was monitored by connecting a multi-meter to the two metal contacts and by moving the sample stage we can spatially map the LITE current. The radius of the laser beam size is ~ 1 μm (see Method) through a confocal microscope in order to get better spatial resolution. When the laser illuminates the left side of the flake, the local temperature of this side will be higher than the right. Therefore, the carrier concentration of the left end will also be higher, which creates a density gradient and the diffusion current flows from left to right, as described in the top panel of Figure 3(b). Similarly, when the laser spot moves to the right, the thermoelectric current will flip the sign. Figure 3(c) shows an optical image of a typical device with Ni contacts and its corresponding LITE current mapping. The thicknesses of the films we used for LITE measurements are in the range of 10-15 nm. In Figure 3(c) the positive sign (red region in the plot) corresponds the current flowing from the left end of device towards the right end and the negative sign (blue) is the other way around, as we expected for p-type semiconductor in Figure 3(b). Also it is worth to mention that we achieved over 3 μA thermoelectric current under 3 mW laser power, which is at least two orders higher (normalized with laser power) than other reported 2D thin film devices such as $SnS_2$[31] and $MoS_2$[30] with similar photo-thermoelectric measurement.

One concern for interpreting LITE current mapping is that photovoltaic effect may mix into the thermoelectric effect and therefore contribute to the total current. Noted that, in a symmetric device structure as in Figure 3c, there is no band tilting in the channel region. The photovoltaic effect only occurs when the laser spot moves to the metal-to-semiconductor interface where the misalignment between the metal work function and semiconductor Fermi level causes the formation a Schottky barrier and introduces electrical field which will collect the photo-generated electron and hole pairs. The electrical field at two contacts will be in the opposite direction, and the photovoltaic current mapping profile may show some resemblance to the thermoelectrical current. One needs to carefully deal with this issue and isolate these two components spatially as discussed by Lee et al. in $SnS_2$ device[31]. The Schottky barrier width (the width of band-bending region), strongly related with carrier concentration, is usually less than 1 μm in heavily doped semiconductors. Hence the photovoltaic current can only be detected within the narrow region at the metal-to-semiconductor interface, whereas the thermoelectric current should be observed through the entire flake with a gradual change from positive to negative value, as shown in Figure 3(c) and 3(d). We can use this difference to spatially separate these two components.

Here we changed metals with different work function to understand the impact of contact metals on photovoltaic effect. There are in general three types of metal-to-semiconductor contacts[34]: accumulation-type, neutral-type (near flat-band condition) and depletion-type, depending on the alignment of metal work function and semiconductor Fermi level, as shown in Figure 4(a), 4(d) and 4(g). In principle by changing the metal work function we should be able to achieve all three types of contacts on the same semiconductor. However, the contact profile is difficult to alter experimentally in most of traditional 3D semiconductors such as Si, Ge and III-V due to the strong Fermi Level pinning effect. For instance, in silicon the contact is usually pinned approximately 1/3 of the bandgap near valence band to form a depletion contact no matter what metal contact is used[35]. It is believed that the dangling bonds at semiconductor interface will introduce high density of mid-bandgap interface states that should be preferentially filled when the metal encounters the semiconductor[34,36]. Therefore, it is difficult to form an accumulation-mode contact (especially in p-type semiconductors[37]) with low resistivity contacts. It is widely believed that the Fermi level pinning issue is less severe in 2D vdW semiconductors than in traditional bulk 3D semiconductors, thanks to their dangling bond-free crystal surface. The pinning factor in 2D materials has even been reported to approach unity, which means no Fermi-level pinning occurs, using transferred contact technique[38]. Here we adopted three different metals: Pd (work function $\Psi_M$ of 5.4 eV), Ni (5.0 eV) and Cr (4.5 eV) for thermoelectric current mapping.

The results of photo-current mapping on three devices with different metal contacts are presented in Figure 4. The Cr-contacted device shows four distinctive areas with different current origin (see Figure 4(d)): the middle red and blue area is thermally generated current and the outer zones are the photovoltaic current generated at the Schottky contacts. By comparing the optical image and the current mapping, we noticed the center of the outer zone is located exactly at the metal-to-semiconductor interface, which confirms the origin of photovoltaic current from the Schottky contacts. According to the polarity of the photovoltaic current, we can deduce that the band at the semiconductor boundary is bending downwards to form a depletion-mode contact (shown in Figure 4(a)), which is the most common case.

In contrast, the Ni-contacted device only have two distinctive thermoelectric current area (Figure 4(b) and 4(e)), indicating there is almost no band bending within our instrument detecting range at the contact region (Figure 4(h)). This corresponds to the flat-band condition which is represented in Figure 4(b). As we adopting higher work function metal Pd, we noticed that the polarity of photovoltaic current has inverted compared to Cr-contacted device (Figure 4(f) and Figure 4(i)), indicating accumulation-type contacts are formed. This suggests Ni and Pd can form near Ohmic metal-to-semiconductor contacts with essentially no Schottky barrier.It explains why the record low contact resistance can be achieved in 2D Te transistors with Pd contacts[22]. It is worth mentioning that such accumulation-mode contacts are very rare cases, especially in p-type semiconductors. The tunable contact band bending profile can be ascribed to the Fermi-level de-pinning effect at van der Waals interfaces. We also noticed that the thermoelectric current is 3 times larger in Pd- and Ni- contacted devices compared to Cr-contacted device, indicating that low resistivity contacts are necessary to collect thermally generated carriers and to harvest the thermoelectric energy more efficiently. By taking advantage of different metal contacts, we can also tune the behavior of 2D Te Schottky transistors (see Figure S2), or implement device structures with versatile functionalities such as Te Schottky diodes with on/off ratio over $10^3$ at room temperature (see Figure S3) and solar cell devices (see Figure S4).

In summary, we present a newly-developed hydrothermally-synthesized versatile 2D Te nano-films as a promising candidate for thermoelectric applications, given its high electrical conductivity, low thermal conductivity and two-dimensionality. Room temperature Seebeck coefficient, power factor and ZT value were measured to be 413 µV/K, 31.7 µW/cm·K$^2$ and 0.63, respectively, on a 31-nm-think 2D Te film. Laser-induced current mapping were performed to visualize the thermoelectric current distribution as well as to understand how the metal-to-semiconductor contacts will impact the efficiency of harvesting thermoelectric power. We observed and identified three distinctive contact types with three different metals. Among them, accumulation-mode contacts were clearly observed in p-type 2D semiconductors for the first time. Our work finds a niche to apply 2D Te film as a potential thermoelectric material for micro-energy harvesting systems or as micro-Peltier coolers. Moreover, the tunable metal-to-semiconductor contacts enable us to design high-performance electronics with novel functionalities.

(1)     Rowe, D. M. *CRC Handbook of Thermoelectrics*; Rowe, D., Ed.; CRC Press, 1995.

(2)     Rowe, D. M. *Thermoelectrics Handbook : Macro to Nano*; CRC/Taylor & Francis, 2006.

(3)     Wright, D. A. Thermoelectric Properties of Bismuth Telluride and Its Alloys. *Nature*. 1958, pp 834–834.

(4)     Polvani, D. A.; Meng, J. F.; Chandra Shekar, N. V.; Sharp, J.; Badding, J. V. Large Improvement in Thermoelectric Properties in Pressure-Tuned p-Type $Sb_{1.5}Bi_{0.5}Te_3$. *Chem. Mater.* **2001**, *13* (6), 2068–2071.

(5)     Venkatasubramanian, R.; Siivola, E.; Colpitts, T.; O'Quinn, B. Thin-Film Thermoelectric Devices with High Room-Temperature Figures of Merit. *Nature* **2001**, *413* (6856), 597–602.

(6)     Jones, W.; March, N. H. *Theoretical Solid State Physics: Perfect Lattices in Equilibrium*; Courier Corporation, 1985; Vol. 1.

(7)     Kittel, C.; McEuen, P.; McEuen, P. *Introduction to Solid State Physics*; Wiley New York, 1996; Vol. 8.

(8)     Bejan, A.; Kraus, A. D. *Heat Transfer Handbook*; John Wiley & Sons, 2003; Vol. 1.

(9)     Dresselhaus, M. S.; Chen, G.; Tang, M. Y.; Yang, R.; Lee, H.; Wang, D.; Ren, Z.; Fleurial, J. P.; Gogna, P. New Directions for Low-Dimensional Thermoelectric Materials. *Adv. Mater.* **2007**, *19* (8), 1043–1053.

(10)   Francis J. DiSalvo. Thermoelectric Cooling and Power Generation. *Science.* **1999**, *285*, 28.

(11)   Pei, Y.; Shi, X.; Lalonde, A.; Wang, H.; Chen, L.; Snyder, G. J. Convergence of Electronic Bands for High Performance Bulk Thermoelectrics. *Nature* **2011**, *473* (7345), 66–69.

(12)   Heremans, J. P.; Dresselhaus, Mildred S.; Bell, L. E.; Morelli, D. T. When

Thermoelectrics Reached the Nanoscale. *J. Pharm. Sci.* **2103**, *8*, 471–473.

(13) Qiu, G.; Wang, Y.; Nie, Y.; Zheng, Y.; Cho, K.; Wu, W.; Ye, P. D. Quantum Transport and Band Structure Evolution under High Magnetic Field in Few-Layer Tellurene. Nano Lett. 2018, 18, 5760–5767.

(14) Zeng, J.; He, X.; Liang, S.-J.; Liu, E.; Sun, Y.; Pan, C.; Wang, Y.; Cao, T.; Liu, X.; Wang, C.; et al. Experimental Identification of Critical Condition for Drastically Enhancing Thermoelectric Power Factor of Two-Dimensional Layered Materials. *Nano Lett.* **2018**, *18* (12), 7538–7545.

(15) Peng, H.; Kioussis, N.; Snyder, G. J. Elemental Tellurium as a Chiral P-Type Thermoelectric Material. *Phys. Rev. B* **2014**, *89* (19), 195206.

(16) Lin, S.; Li, W.; Chen, Z.; Shen, J.; Ge, B.; Pei, Y. Tellurium as a High-Performance Elemental Thermoelectric. *Nat. Commun.* **2016**, *7*, 10287.

(17) Nakayama, K.; Kuno, M.; Yamauchi, K.; Souma, S.; Sugawara, K.; Oguchi, T.; Sato, T.; Takahashi, T. Band Splitting and Weyl Nodes in Trigonal Tellurium Studied by Angle-Resolved Photoemission Spectroscopy and Density Functional Theory. *Phys. Rev. B* **2017**, *95* (12), 1–5.

(18) Agapito, L. A.; Kioussis, N.; Goddard, W. A.; Ong, N. P. Novel Family of Chiral-Based Topological Insulators: Elemental Tellurium under Strain. *Phys. Rev. Lett.* **2013**, *110* (17), 1–5.

(19) Sharma, S.; Singh, N.; Schwingenschlögl, U. Two-Dimensional Tellurene as Excellent Thermoelectric Material. *ACS Appl. Energy Mater.* **2018**, *1* (5), 1950–1954.

(20) Lin, C.; Cheng, W.; Chai, G.; Zhang, H. Thermoelectric Properties of Two-Dimensional Selenene and Tellurene from Group-VI Elements. *Phys. Chem. Chem. Phys.* **2018**, *06*, 1.

(21) Gao, Z.; Tao, F.; Ren, J. Unusually Low Thermal Conductivity of Atomically Thin 2D Tellurium. *Nanoscale* **2018**, *10* (27), 12997–13003.

(22) Wang, Y.; Qiu, G.; Wang, R.; Huang, S.; Wang, Q.; Liu, Y.; Du, Y.; Goddard, W. A.; Kim, M. J.; Xu, X.; et al. Field-Effect Transistors Made from Solution-Grown Two-Dimensional Tellurene. *Nat. Electron.* **2018**, *1* (4), 228–236.

(23) Du, Y.; Qiu, G.; Wang, Y.; Si, M.; Xu, X.; Wu, W.; Ye, P. D. One-Dimensional van Der Waals Material Tellurium: Raman Spectroscopy under Strain and Magneto-Transport. *Nano Lett.* **2017**, *17*, 3965−3973.

(24) Qiu, G.; Si, M.; Wang, Y.; Lyu, X.; Wu, W.; Ye, P. D. High-Performance Few-Layer Tellurium CMOS Devices Enabled by Atomic Layer Deposited Dielectric Doping Technique. *2018 76th Device Res. Conf.* **2018**, *06202* (2017), 1–2.

(25) Wu, W.; Qiu, G.; Wang, Y.; Wang, R.; Ye, P. Tellurene: Its Physical Properties, Scalable Nanomanufacturing, and Device Applications. *Chem. Soc. Rev.* **2018**, *47* (19), 7203–7212.

(26) Gao, S.; Wang, Y.; Wang, R.; Wu, W. Piezotronic Effect in 1D van Der Waals Solid of Elemental Tellurium Nanobelt for Smart Adaptive Electronics. *Semicond. Sci. Technol.* **2017**, *32* (10).

(27) Amani, M.; Tan, C.; Zhang, G.; Zhao, C.; Bullock, J.; Song, X.; Kim, H.; Shrestha, V. R.; Gao, Y.; Crozier, K. B.; et al. Solution-Synthesized High-Mobility Tellurium Nanoflakes for Short-Wave Infrared Photodetectors. *ACS Nano* **2018**, *12* (7), 7253–7263.

(28) Gehring, P.; Harzheim, A.; Spiece, J.; Sheng, Y.; Rogers, G.; Evangeli, C.; Mishra, A.; Robinson, B. J.; Porfyrakis, K.; Warner, J. H.; et al. Field-Effect Control of Graphene-Fullerene Thermoelectric Nanodevices. *Nano Lett.* **2017**, acs.nanolett.7b03736.

(29) Kodama, T.; Ohnishi, M.; Park, W.; Shiga, T.; Park, J.; Shimada, T.; Shinohara, H.; Shiomi, J.; Goodson, K. E. Modulation of Thermal and Thermoelectric Transport in

Individual Carbon Nanotubes by Fullerene Encapsulation. *Nat. Mater.* **2017**, *16* (September).

(30) Kayyalha, M.; Maassen, J.; Lundstrom, M.; Shi, L.; Chen, Y. P. Gate-Tunable and Thickness-Dependent Electronic and Thermoelectric Transport in Few-Layer $MoS_2$. *J. Appl. Phys.* **2016**, *120* (13).

(31) Lee, M.-J.; Ahn, J.-H.; Sung, J. H.; Heo, H.; Jeon, S. G.; Lee, W.; Song, J. Y.; Hong, K.-H.; Choi, B.; Lee, S.-H.; et al. Thermoelectric Materials by Using Two-Dimensional Materials with Negative Correlation between Electrical and Thermal Conductivity. *Nat. Commun.* **2016**, *7*, 12011.

(32) Luo, Z.; Maassen, J.; Deng, Y.; Du, Y.; Lundstrom, M. S.; Ye, P. D.; Xu, X. Anisotropic In-Plane Thermal Conductivity Observed in Few-Layer Black Phosphorus. *Nat. Commun.* **2015**, *6*, 1–32.

(33) Ho, C. Y.; Powell, R. W.; Liley, P. E. Thermal Conductivity of the Elements. *J. Phys. Chem. Ref. Data* **1972**, *1* (2), 279–421.

(34) Sze, S. M.; Ng, K. K. *Physics of Semiconductor Devices*; John wiley & sons, 2006.

(35) Allen, F. G.; Gobeli, G. W. Work Function, Photoelectric Threshold, and Surface States of Atomically Clean Silicon. *Phys. Rev.* **1962**, *127* (1), 150.

(36) Spicer, W. E.; Chye, P. W.; Garner, C. M.; Lindau, I.; Pianetta, P. The Surface Electronic Structure of III-V Compounds and the Mechanism of Fermi Level Pinning by Oxygen (Passivation) and Metals (Schottky Barriers). *Surf. Sci.* **1979**, *86* (C), 763–788.

(37) Nishimura, T.; Kita, K.; Toriumi, A. Evidence for Strong Fermi-Level Pinning Due to Metal-Induced Gap States at Metal/Germanium Interface. *Appl. Phys. Lett.* **2007**, *91* (12).

(38) Liu, Y.; Guo, J.; Zhu, E.; Liao, L.; Lee, S.-J.; Ding, M.; Shakir, I.; Gambin, V.; Huang, Y.;

Duan, X. Approaching the Schottky–Mott Limit in van Der Waals Metal–semiconductor Junctions. *Nature* **2018**, *557* (7707), 696–700.

## Method

### Device fabrication

The 2D Te flakes were synthesized and transferred onto Silicon wafer with 90 nm $SiO_2$ capping layer following the procedure reported in ref. 22. The devices were patterned with electron beam lithography. 10 nm Ni and 100 nm Pt were deposited with electron beam evaporator as two thermometers since Pt has better linearity in temperature coefficient. 10 nm Ni and 100 nm Au were deposited for the rest of the contacts.

The devices for current mapping experiment were fabricated with the similar procedure, expect that three different types of metals were deposited (30/100 nm Cr/Au, 30/100 nm Ni/Au, 30/100 nm Pd/Au).

### Seebeck coefficient and power factor measurement

A heating power was added to the system by applying a DC voltage (from 1V to 10V) across the heater with Keithley 2450 voltage source meter. After sufficiently long time, the system reaches equilibrium, and the voltage drop and temperature gradient stabilized. The voltage difference across the flake was measured with Keithley 2182A nanovoltmeter. The resistance of two thermometers were measured with synchronized Keithley 6221 current source meter and Keithley 2182A nanovoltmeter in delta mode and the local temperature were determined by converting the resistance of the thermometer into temperature through pre-calibrated temperature versus resistance curve. The electrical conductivity of Te films were measured in four-terminal configuration using SR830 lock-in amplifier with standard AC measurement techniques.

### Laser induced current mapping:

The photocurrent is driven by a normal incident He-Ne laser (633-nm). The light is focused by a 50x long working-distance objective (Nikon CF Plan 50x EPI SLWD, $N_A$0.45) to a $w_0 = 0.96$ μm spot (measured by standard knife-edge method). A Keithley 2612a source meter is coupled with a piezoelectric nano-positioner (MadCityLab NanoLP-100) to conduct the spatial mapping of current measurement. The photocurrent is measured at zero bias $V_{ds} = 0$ V.

**Figures**

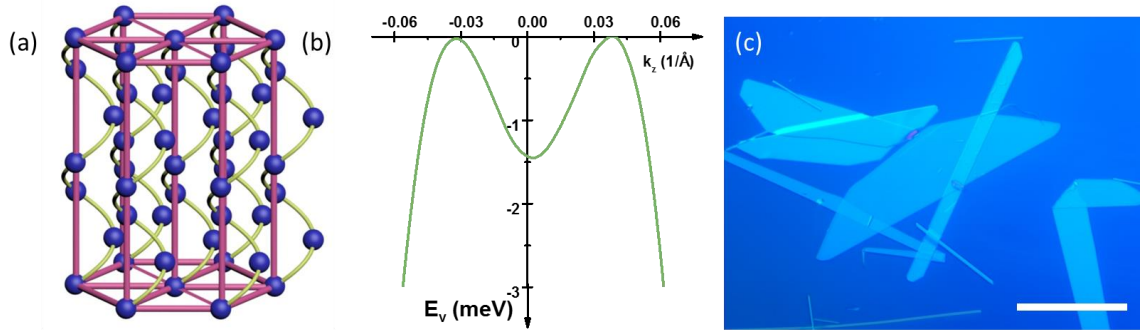

**Figure 1| Atomic structure, band structure, and 2D Te nanofilms.** (a) One-dimensional van der Waals chiral structure of Te crystal. The nearest neighboring chiral chains are bonded by vdW forces into a hexagonal structure. (b) The band structure in the vicinity of valence band maxima at H point. (c) Optical image of as-grown 2D Te films dispensed onto a silicon wafer. The scale bar is 50 μm. The 1D helical chains are aligned with the long edge of Te flakes.

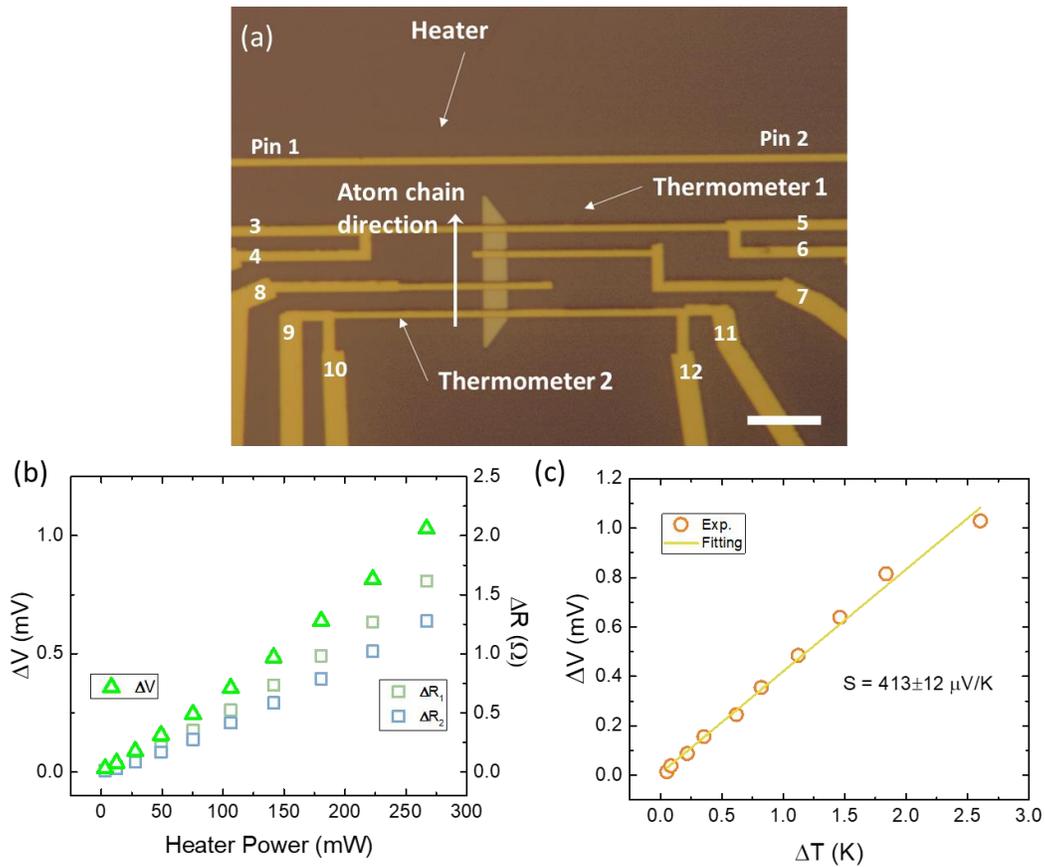

**Figure 2| Seebeck coefficient measurement of 2D Te.** (a) Device structure and pin-out diagram. The scale bar is 10 μm. Pin 1 and 2 are connected to the micro-heater that generates a temperature gradient. Pin3-6 and Pin 9-12 are two thermometers that measure the local temperature. Combining Pin 3, 7, 8 and 9 we can also measure the conductivity of the Te film with four-terminal configuration. (b) The voltage drop across the film and the resistance change of two thermometers as functions of heater output power. (c) The extraction of Seebeck coefficient. The voltage drop and the temperature gradient shows linear relationship where the slope give the Seebeck coefficient $S = 413 \pm 12 \ \mu V/K$.

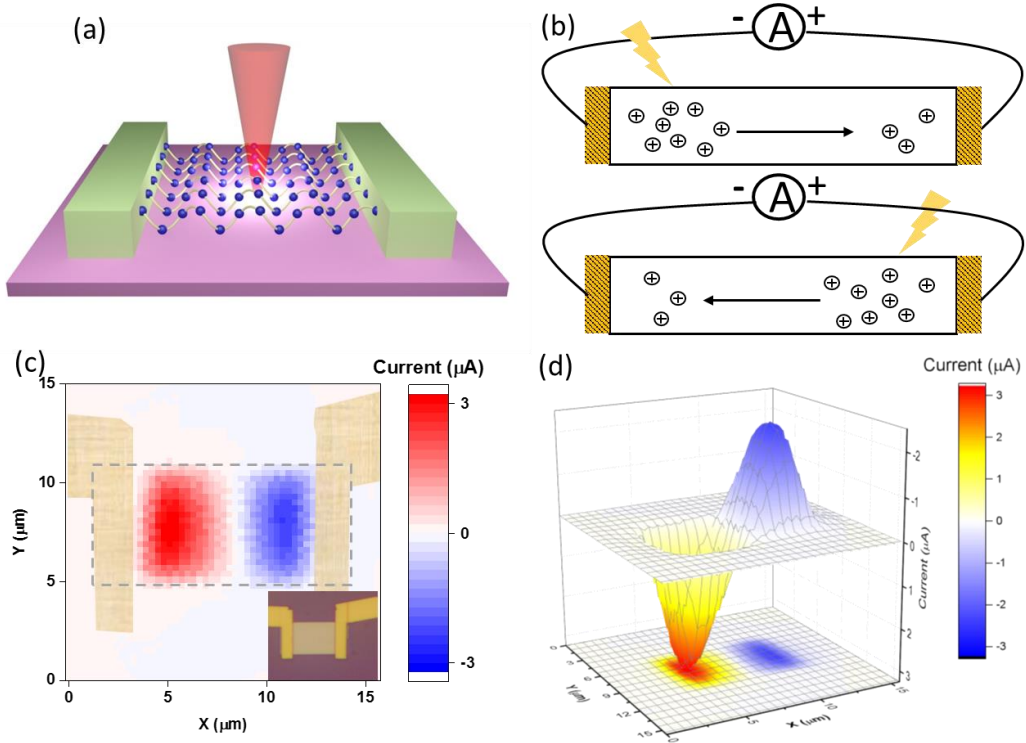

**Figure 3| Laser Induced Thermoelectric (LITE) current mapping.** (a) Schematics of the experiment setup. Two Ni contacts were patterned onto the flake. A 633 nm He-Ne laser was used to heat up the flake locally. (b) The principle of LITE mapping. For p-type devices, when the laser illuminates the left side of the flake (top panel), the carrier concentration of the left will be higher than right side which induce a diffusion current from the left to right inside the flake. When the laser shines on the right side the current direction will be flipped (bottom panel). (c) The LIFE current mapping of the real device. The inset is the optical image of the same device. (d) The LITE current mapping in a 3D plot.

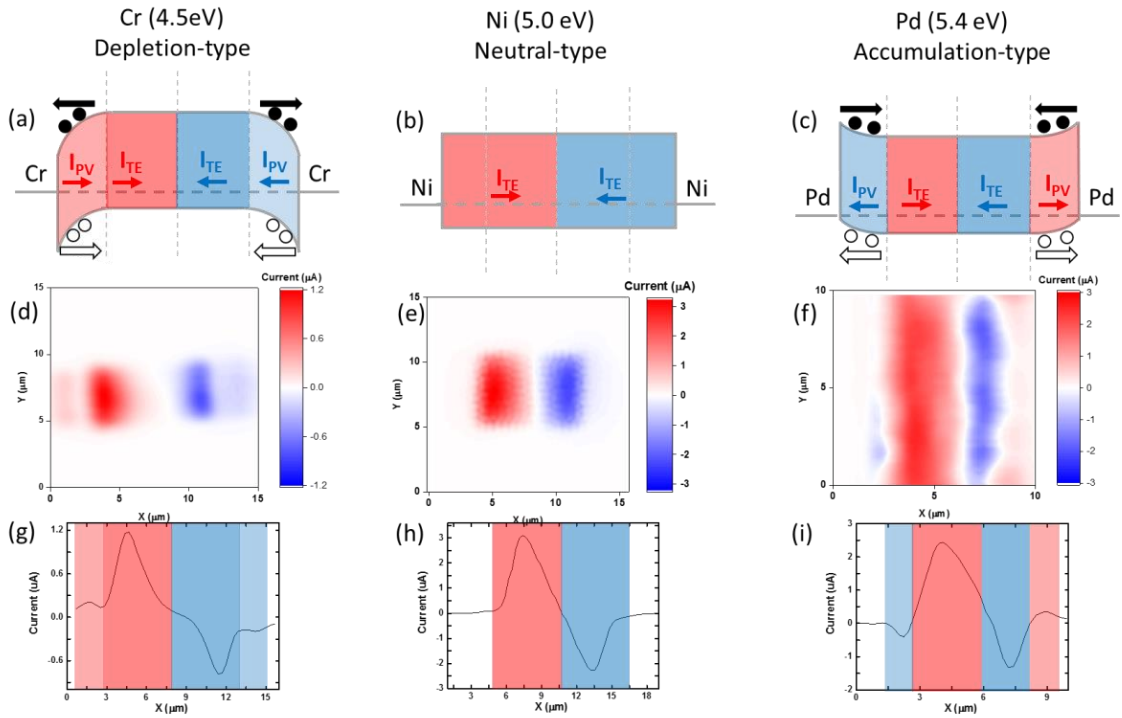

**Figure 4| Visualizing three types of metal-to-semiconductor contacts by laser induced current mapping.** (a)-(c) Band diagrams of depletion-mode (Cr-contacted), neutral-mode (Ni-contacted) and accumulation-mode (Pd-contacted) devices. The laser induced current mapping will be the superposition of thermoelectric current ($I_{TE}$) in the middle region and photovoltaic current ($I_{PV}$) at the interface. (d) - (f) Current mapping of three devices with Cr, Ni and Pd contacts. The red and blue color correspond to the current flow to the right and to the left, respectively, the same as in (a) – (c). (g) - (i) The line profile of the current distribution cut along the center of the channel in three devices.

Supplementary Information for:

# Thermoelectric Performance of 2D Tellurium with Accumulation Contacts

Gang Qiu[1, 2], Shouyuan Huang[2, 3], Mauricio Segovia[2,3], Prabhu K. Venuthurumilli[2, 3], Yixiu Wang[4], Wenzhuo Wu[4], Xianfan Xu[1,2, 3]⋆, Peide D. Ye[1, 2]⋆,

[1]School of Electrical and Computer Engineering, Purdue University, West Lafayette, Indiana 47907, USA

[2]Birck Nanotechnology Center, Purdue University, West Lafayette, Indiana 47907, USA

[3]School of Mechanical Engineering, Purdue University, West Lafayette, Indiana 47907, USA

[4]School of Industrial Engineering, Purdue University, West Lafayette, Indiana 47907, USA

⋆e-mail: Correspondence and requests for materials should be addressed to X.X. (xxu@ecn.purdue.edu) or P. D. Y. (yep@purdue.edu)

**Supplementary Note 1: Temperature coefficient calibration**

In order to convert the resistance of the Pt/Ni nano-strip thermometer, the temperature coefficient was first calibrated for a broad temperature range. The resistance was measured with lock-in amplifier using standard AC method. The sample was cooled down in a dilution refrigerator from room temperature to helium temperature. The temperature versus resistance curve is shown in Figure S1, which exhibits great linearity for a broad temperature range from 300 K to 30 K. The temperature coefficient, defined by the following equation: $\frac{dR}{R} = \alpha T$, is extracted to be 0.0019 at the vicinity of room temperature, which can be used to convert the resistance of the thermometer into temperature.

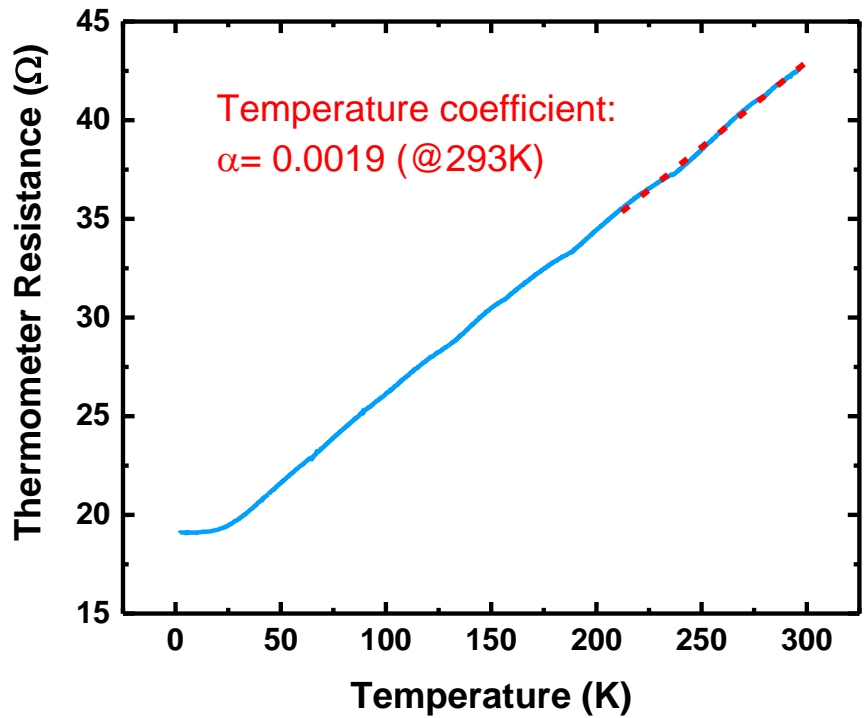

**Figure S1| Calibrating the temperature coefficient of Pt thermometers.** The resistance versus temperature curve shows good linearity for a broad temperature range. The temperature coefficient is extracted from the equation: $\frac{dR}{R} = \alpha T$ to be 0.0019 at the vicinity of room temperature.

**Supplementary Note 2: Tuning the transistor performance with different metal contacts**

To understand how the work function of contact metal will affect the transistor performance, we fabricated multiple transistors with different contact metals on the same flake to eliminate the flake-to-flake variation. The device schematics was presented in Figure S2(a). A 2D Te flake was transferred onto a heavily doped silicon wafer as back gate with 90 nm $SiO_2$ insulating layer. Two high work function metal (Pd, work function ~5.2eV) and two low work function metal (Cr, work function ~4.6 eV) was patterned with even spacing and deposited with electron beam evaporator. We now have a Pd-Te-Pd transistor, a Cr-Te-Cr transistor, and a Pd-Te-Cr Schottky diode. The transfer curves of these three devices were measured with the modulation of the back gate and plotted in the same figure in Figure S3(b). We notice that the Pd-contacted device shows strongest p-type behavior and largest drain current since the work function is aligned close to the valence band and there are no obstacles for hole transport. The Cr-Cr shows ambipolar behavior and the drain current is much smaller compared to Pd device since both the electron and hole current branch suffered from the large Schottky barrier. The Pd-Cr Schottky current shows asymmetric device performance: when device was forward biased, the hole current can be injected through Pd contact smoothly whereas when the device was reversely biased the current will be hindered by Cr contact.

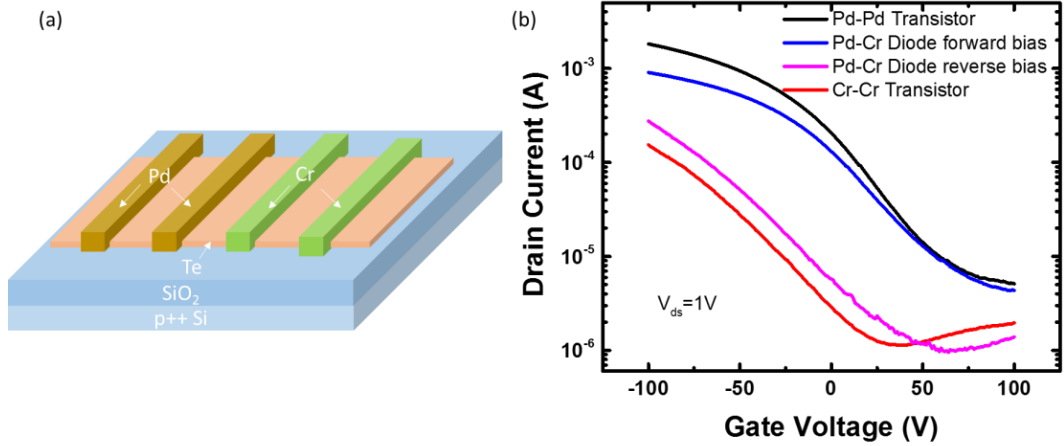

**Figure S2| Multiple channel transistors with different metal contacts on the same flake**. (a) Schematic image of device structure. Two Pd contacts and two Cr contacts were placed on the same Te flake with even spacing. (b) Transfer curves of Pd-Pd transistor, Cr-Cr transistor, and Pd-Cr diode with forward and reverse biases.

**Supplementary Note 3: Te Schottky diodes**

Owing to the controllability of Te-to-metal contact behavior, we can achieve high-performance Te Schottky diodes. Figure S3 shows a Te Schottky diodes with Pd and Ti contacts respectively. A large rectification ratio over $10^3$ was measured at room temperature.

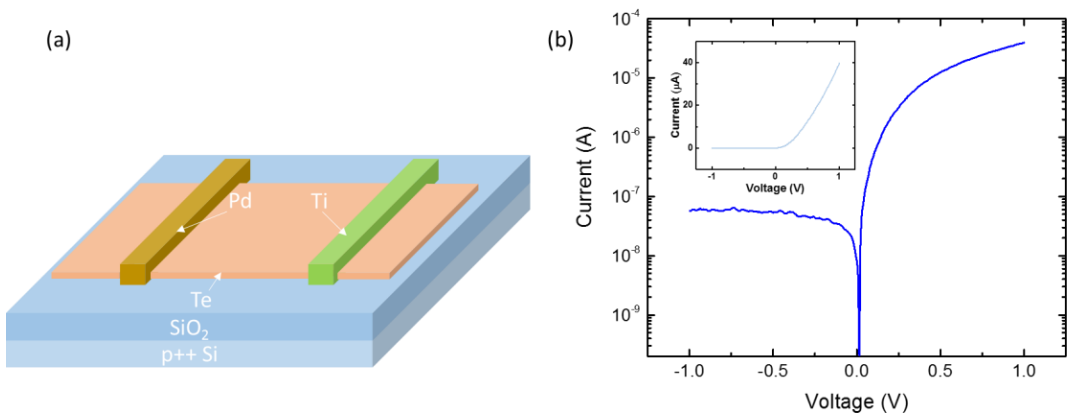

**Figure S3| Te-based Schottky diode devices.** (a) Schematic image of Schottky device structure. Pd and Ti contacts were fabricated onto 2D Te flakes. (b) I-V curves of a Schottky diode plotted in log scale. The rectification ratio is over $10^3$. Inset: the I-V current plotted in linear scale.

**Supplementary Note 4: Photovoltaic measurement based on Te Schottky diodes**

The Schottky diodes can be used as solar cells. Here we also explore the photovoltaic effect of Te Schottky devices. The Pd-Te-Cr diode devices with channel length of 5 μm were fabricated. 633 nm laser illuminated the center of the flake and I-V curves with and without the laser was plotted in the inset of Figure S4(a). Particularly for solar cells, we are interested in two figures-of-merit, the short-circuit current and open-circuit voltage which can be determined from the intercept of I-V curve, as shown in Figure S4(a). The output power of the solar cell $P_d$ can be calculated by multiplying the current and voltage which peaks at ~ 0.25 nW at 0.7 mV, as shown in Figure S4(b).

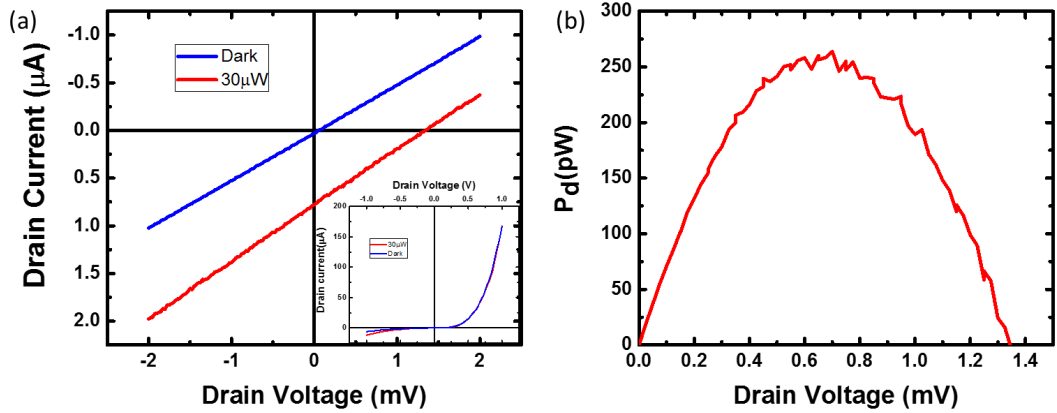

**Figure S4| Demonstration of solar cell device based on Te Schottky diodes.** (a) I-V curve of a Pd-Cr diodes with and without laser illumination. The open-circuit voltage and short-circuit current can be extracted from the intercept of the red I-V curve. Inset, the I-V curve on a broader sweep range. (b) The output power of the solar cell $P_d$ can be calculated by multiplying the current and voltage which peaks at ~ 0.25 nW at 0.7 mV.

**Supplementary Note 5: In-plane thermal conductivity measured using Micro-Raman method**

The thin-film tellurium sample (35-nm) for thermal conductivity measurement is suspended on a silicon trench (5-µm um wide, 13-µm deep) by the PDMS stamping method[1] using the Langmuir–Blodgett transfer process[2]. A 633-nm He-Ne laser beam is diffracted through a slit, and then focused using a 100x objective into a laser focal line. The length of the laser focal line is 7.5 µm, and is along the direction of the silicon trench, and the width is 0.45 µm, positioned at the center of the suspending region. The laser focal line heats up the tellurium film, creating a temperature gradient and inducing Raman scattering at the same time. The Raman scattering signal is collected and sent to a spectrometer (Horiba LabRam). By calibrating the Raman peak shift with temperature, the Raman signal can be used as a thermometer. Laser absorption by the tellurium film is determined by measuring reflection and transmission. The thermal conductivity can then be extracted using a heat transfer model with the knowledge of laser absorption, Raman shift/temperature, and geometry. This method is adapted from our previous works for measuring in-plane thermal conductivity of 2D materials[3–5].

Figure S5(a) shows the measured Raman spectrum of the suspended 35-nm tellurium thin-film. The Raman peak shift of $A_1$ mode vs. temperature is shown in Figure S5(b) and is used as the thermometer, due to its high temperature resolution and good linearity. The sample is then heated using the laser focal line, and the Raman shift vs. laser power is showed in Figure S5(c). Comparing to the numerical model (Figure S5(d)), the thermal conductivity is extracted to be 1.50 W/m-K. The detailed studies on thermal transport of Te films including numerical calculations are being prepared for another publication.

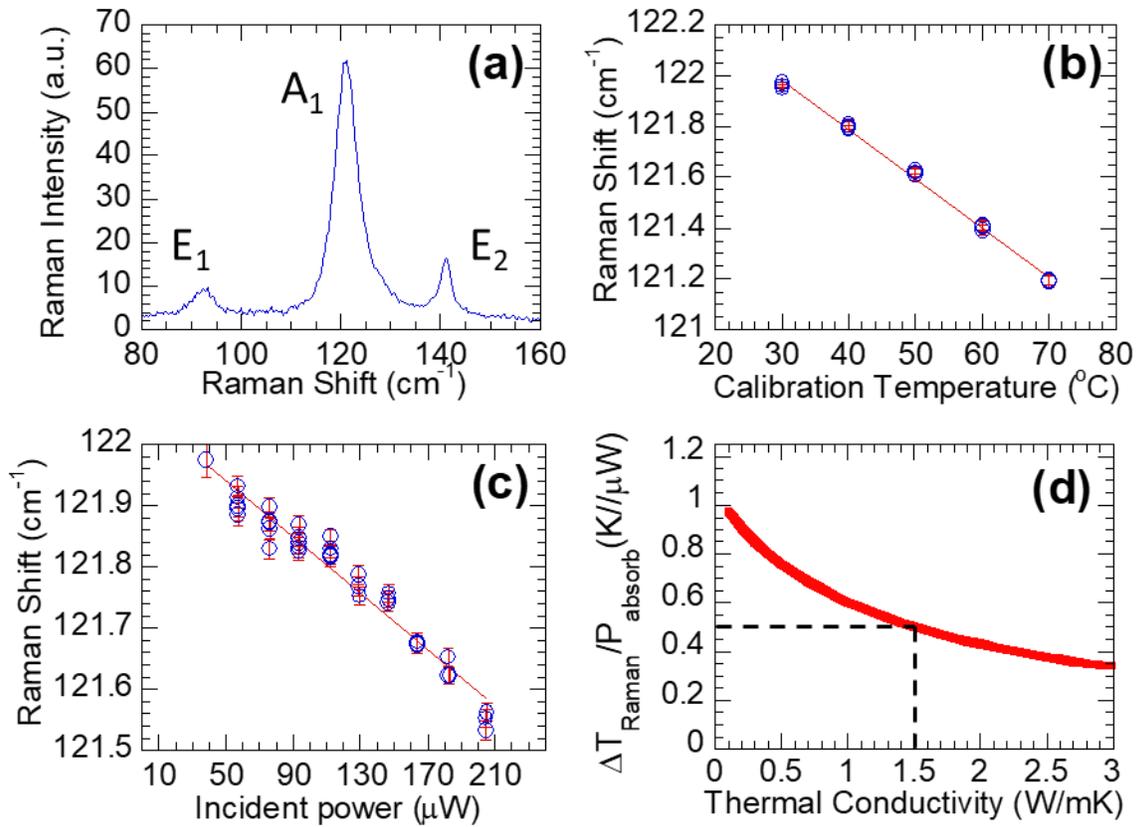

**Figure S5| Measurement of in-plane thermal conductivity using micro-Raman thermometry.** (a) Typical Raman spectrum of thin-film tellurium. (b) Calibration of peak position of $A_1$ mode vs. temperature as a thermometer. (c) $A_1$ mode Raman shift vs. incident laser power. (d) Numerical modeling result of temperature rise per absorbed laser power. Dashed lines indicate the in-plane thermal conductivity of 1.50 W/m-K, corresponding to the measurement temperature rise per absorbed laser power of 0.503 K/µW.

(1) Ma, X.; Liu, Q.; Xu, D.; Zhu, Y.; Kim, S.; Cui, Y.; Zhong, L.; Liu, M. *Nano Lett.* **2017**, *17* (11), 6961–6967.

(2) Wang, Y.; Qiu, G.; Wang, R.; Huang, S.; Wang, Q.; Liu, Y.; Du, Y.; Goddard, W. A.; Kim, M. J.; Xu, X.; Ye, P. D.; Wu, W. *Nat. Electron.* **2018**, *1* (4), 228–236.

(3) Luo, Z.; Liu, H.; Spann, B. T.; Feng, Y.; Ye, P.; Chen, Y. P.; Xu, X. *Nanoscale Microscale Thermophys. Eng.* **2014**, *18* (2), 183–193.

(4) Luo, Z.; Maassen, J.; Deng, Y.; Du, Y.; Lundstrom, M. S.; Ye, P. D.; Xu, X. *Nat. Commun.* **2015**, *6*, 1–32.

(5) Luo, Z.; Tian, J.; Huang, S.; Srinivasan, M.; Maassen, J.; Chen, Y. P.; Xu, X. *ACS Nano* **2018**, *12* (2), 1120–1127.